\documentclass[12pt,]{article}
\usepackage{amsfonts}
\usepackage{amsmath,amssymb,bm}
\usepackage{graphicx}

\begin{document}

\title{On the effects of conformal degrees of freedom inside a neutron
star}
\author{F. Canfora, A. Giacomini, S. Willison. \\
{\small \textit{Centro de Estudios Cientificos (CECS), Casilla 1469
Valdivia, Chile.}}\\
{\small e-mails: \textit{canfora@cecs.cl, giacomini@cecs.cl, steve@cecs.cl.}}%
}
\date{}
\maketitle

\begin{abstract}
In this paper a neutron star with an inner core which undergoes a
phase transition, which is characterized by conformal degrees of
freedom on the phase boundary, is considered. Typical cases of such
a phase transition are e.g. quantum Hall effect, superconductivity
and superfluidity. Assuming the mechanical stability of this system
the effects induced by the conformal degrees of freedom on the phase
boundary will be analyzed. We will see that the inclusion of
conformal degrees of freedom is not always consistent with the
staticity of the phase boundary. Indeed also in the case of
mechanical equilibrium there may be the tendency of one phase to
swallow the other. Such a shift of the phase boundary would not
imply any compression or decompression of the core. By solving the
Israel junction conditions for the conformal matter, we have found
the range of physical parameters which can guarantee a stable
equilibrium of the phase boundary of the neutron star. The relevant
parameters turn out to be not only the density difference but also
the difference of the slope of the density profiles of the two
phases. The values of the parameters which guarantee the stability
turn out to be in a phenomenologically reasonable range. For the
parameter values where the the phase boundary tends to move, a
possible astrophysical consequence related to sudden small changes
of the moment of inertia of the star is briefly discussed.
\end{abstract}


Preprint: CECS-PHY-07/20 \newline

\section{Introduction}

One of the most intriguing features of neutron stars is the
possibility of having an inner core which undergoes a (colour)
superconductivity and/or superfluidity phase transition due to the
extremely high pressure and density. This fact was first pointed out
by Migdal \cite{Mi59} (there is a huge amount of literature on this
subject with a little hope of providing one
with a complete list of references: for two detailed reviews, see\ \cite%
{Pet92} \cite{Ra99} and references therein; for general relativistic
formalisms suitable for dealing with neutron stars within these
scenarios see \cite{CL98} \cite{AC07} and references therein). Such
phase transitions inside neutron stars could lead to interesting
observational effects relevant in cosmology \cite{Sigl:2006ur} and
through the quasi normal modes of the stars as well as the related
gravitational radiation (see, for
instance, \cite{ST02} \cite{MP03}; detailed reviews on the subjects
are \cite%
{KS99} \cite{Ste03} \cite{ST03}). Recently, there has been also
pointed out the intriguing possibility to have a quantum hall phase
of gluons and quarks in
the inner core of a neutron star (see, in particular, \cite{IM05}
\cite{Iw05}%
). The discovery of \textit{magnetars} \cite{DT92}, highly magnetized
neutron stars whose magnetic fields can reach $10^{18}G$ in the core, makes
the arising of Quantum Hall features in the inner core of a neutron star not
unlikely. All the here described types of phase transitions have in common
to be characterized by the presence of conformal degrees of freedom on the
phase boundary predicted by QFT\footnote{%
The fact that the boundary degrees of freedom are conformal can be
seen "heuristically" in the case of a superconducting phase. Due to
the Meissner effect, the current must circulate on the phase
boundary of the superconductive phase. Phenomenologically, there is
no dissipation, this means that there is no physical scale over
which the boundary excitations can dissipate energy. The lack of a
characteristic scale for the boundary theory is related to conformal
symmetry. In the case of Quantum Hall Effect, this can be easily
seen from the fact that it is well described by a Chern-Simons
action in the bulk. Thus, the theory induced on the boundary is the
Wess-Zumino-Witten action which also has conformal symmetry.} (see,
for instance, \cite{Wi90} \cite{We96}).

The main goal of this paper is to study under which conditions the
phase boundary inside a neutron star, which undergoes such a phase
transition with conformal boundary degrees of freedom, can be static
in general relativity, assuming mechanical equilibrium of the
system\footnote{ The analysis of the mechanical stability of neutron
stars has been performed in a huge number of papers.  To provide one
with a complete list of references is a completely hopeless
task; representative papers are, for instance, \cite{EKO91} \cite%
{ABR00} \cite{CL98} and references therein)}. Indeed also if the
system is mechanically stable i.e. there is no pressure inducing the
collapse or expansion of the core, there can be another sort of
instability: there may be the tendency of one phase to prevail over
the other near the boundary. This would imply that the boundary gets
shifted. This situation is analogous to a bubble chamber where a
small perturbation induces a phase transition. To study this
phenomenon  one must take into account the contribution of the
nontrivial traceless boundary stress tensor to the Einstein field
equations. This problem is equivalent to solving Israel's junction
conditions. It is necessary to assume a dynamical phase boundary
and to check under which conditions such a configuration is a
solution of the Einstein field equations (by solving the junction
conditions). It turns out that the equation of motion for the
dynamical phase boundary is equivalent to the dynamics of a
classical point particle in an effective potential. There exist a
static stable equilibrium configuration if the effective potential
has a local minimum for a suitable negative value of the potential.
It is shown that the existence of a local minimum is regulated both
by the difference in the mass density between the two phases and by
difference of the slope of the two density profiles. A third
parameter which determines the form of the potential is the energy
of the shell. The non trivial effects of conformal boundary degrees
of
freedom have been pointed out, in the context of black hole physics, in \cite%
{Cao7} in which they are related to the arising (after the event horizon is
formed) of Quantum Hall features \cite{Lau99} \cite{Wi90}\ related to the
strong attractive nature of the gravitational field acting on Fermions
inside a collapsed neutron star.

The structure of the paper is the following: in Section 2, the
assumptions of the present paper are explained. In Section 3 the
junction conditions derived from the Einstein equations for a
neutron star with a phase boundary are solved. In section 4 the
range of the physical parameters, characterizing our configuration,
for which the phase boundary is in stable equilibrium is determined.
In section 5 an astrophysical implication is discussed. Section 6
the conclusions and perspectives are drawn.

\section{The standard approximations in a neutron star}

Inside a neutron star, at the energy scale of the standard model of
particle physics (up to $TeV$) the collapsing neutron star can be
described very well by QFT and classical general relativity. The
quantum dynamics of the neutrons and of the quarks living inside the
neutron star is much faster than the dynamics of the gravitational
field. This implies that one can compute the equation(s) of state of
the Fermions as usual and then use such equation(s) to solve the
Einstein equations in which the source is described by the equation
of state itself. The success of this theory initiated by Landau,
Chandrasekhar, Tolman, Oppenheimer, Volkoff, Snyder (and many
others) tells that such an adiabatic approximation is excellent. The
typical mass $M_{NS}$, radius $R_{NS}$, density $\rho _{NS}$ and the
baryon
number $N_{NS}$ a neutron star are%
\begin{eqnarray*}
M_{NS} &\approx &10^{33}g,\qquad R_{NS}\approx \left( 10^{6}\div
10^{7}\right) cm,\text{ } \\
\quad \rho _{NS} &\approx &\left( 10^{11}\div 10^{15}\right)
g/cm^{3},\qquad N_{NS}\approx 10^{54}.
\end{eqnarray*}

A neutron star pulsation can be thought of as a linearized solution
of the Einstein equations around a background describing a static
star (see, for the spherically symmetric case, Eq. (\ref{ns1})) with
a "quasi periodic" time dependence: that is, the perturbation
$h_{\mu \nu }$ of the background metric Eq. (\ref{ns1}) has the
following form
\begin{equation*}
h_{\mu \nu }\left( \overrightarrow{x},t\right) =\exp \left( i\left( \omega +%
\frac{i}{\tau }\right) t\right) h_{\mu \nu }^{\omega ,\tau }\left(
\overrightarrow{x}\right)
\end{equation*}
where the imaginary part of the frequency $\frac{1}{\tau }$ is due to the
damping by emission of gravitational and electromagnetic waves (see, for
instance, \cite{KS99}). The standard order of magnitude of the frequency of
a neutron star pulsation is%
\begin{equation*}
19,85\mathrm{Hz}\lessapprox \omega \lessapprox 12,84\mathrm{kHz}
\end{equation*}%
while the damping has a wider range of possible values varying from low
damping (in which $\tau $ can be of the order of years) to high damping (in
which $\tau $ is of the order of millisecond).

\section{Junction conditions and equation of motion of the phase boundary}

In this section the junction conditions for the considered neutron star are
found and the physical parameters that determine the existence of stable
equilibrium configurations are found. We will set
\begin{equation*}
\hbar =1,\qquad c=1,\
\end{equation*}%
while keeping explicitly the Newton constant.

In the present case, one would like to describe a star in which the
inner core underwent a gapped phase transition and therefore
conformal boundary degrees of freedom localized at the separation
between the two phases (predicted by QFT \cite{Wi90} \cite{We96})
have also to be included. We will therefore assume that there are
two phases, each in mechanical equilibrium, namely the interior
region which represents the exotic (superfluid, superconducting
quantum Hall) phase and the exterior region which is the normal
phase. Each phase is approximated by a perfect fluid in a static
configuration. The two regions are characterised by very different
equations of state.

We shall see that, even assuming static equilibrium (without any
compression or decompression of the core), the conformally invariant
matter living on the phase boundary may not be consistent with a
static junction. In this case, since we assume that the conformal
invariance is a defining characteristic of the phase transition,
there must be some tendency of the phase boundary to move. Such a
change in the position of the phase boundary is not by collapse of
matter but by a kind of ``creeping" phase transition, where one
phase tends to swallow the other.

The metric describing a spherically symmetric neutron star can be
parametrized as follows
\begin{align}
ds^{2} &=-\exp (2\nu )dt^{2}+\exp (2\lambda )dr^{2}+r^{2}d\Omega
^{2}\, ,
\label{ns1} \\
\nu &=\nu (r),\ \ \ \lambda =\lambda (r)  \notag
\end{align}%
where $d\Omega ^{2}$\ is\ the line element of a unit sphere\footnote{%
The description of a rotating neutron star is much more difficult. However,
at least for slowly rotating neutron stars, one can argue that the analytic
results derived here in the spherically symmetric case do not change
qualitatively as long as the "rotation" can be dealt as a perturbation \cite%
{Ste03} \cite{FG07}.}. The functions $\nu $ and $\lambda $ are
determined by the Einstein equations. If the source is a perfect
fluid one gets the
standard Tolman-Oppenheimer-Volkoff-Snyder equations:%
\begin{eqnarray}
\exp (2\lambda ) &=&\left( 1-2G\frac{m(r)}{r}\right) ^{-1},\ \ \ m(r)=4\pi
\int^{r}x^{2}\rho (x)dx  \label{0matov} \\
\partial _{r}P &=&-\left( \rho +P\right) \partial _{r}\nu ,\ \ \ \ \partial
_{r}\nu =\exp (2\lambda )\frac{\left( m+4\pi r^{3}P\right) }{r^{2}}
\label{nsTOV}
\end{eqnarray}%
where $\rho $ and $P$ are the density and the pressure of the perfect fluid.

One has to find under which conditions such a configuration solves
the Einstein field equations. Given the static perfect fluid
solutions in the interior and exterior, this problem is equivalent
to solving Israel's junction conditions \cite{Is66}. This technique
allows the construction of an exact solution of the Einstein
equations by matching two different solutions provided that the
metric is continuous and the discontinuity of the
extrinsic curvature is compensated by a suitable energy momentum tensor $%
S_{\mu }^{\nu }$ describing the boundary degrees of freedom of the
inner phase: in the present case in which the inner phase transition
is gapped one should only allow traceless $S_{\mu }^{\nu }$. Let
$\Sigma $ be a timelike hypersurface of codimension one, on which
the matching has to be performed. Let $n ^{\mu }$ be the (spacelike)
unit normal to $\Sigma $, and let $h^{\mu \nu }=g^{\mu \nu }- n^{\mu
}n^{\nu }$ be the metric induced on $\Sigma$. The matching
conditions at $\Sigma$ are: that the induced metric be continuous;
that the jump in the extrinsic curvature is related to the intrinsic
stress tensor localized on $\Sigma$ by \cite{Is66}
\begin{equation}
\gamma _{\nu }^{\mu }=-8\pi G(S_{\nu }^{\mu }-\frac{1}{2}\delta
_{\nu }^{\mu }trS),  \label{jun1}
\end{equation}%
where $\gamma _{\mu \nu }$ is defined as
\begin{equation*}
\gamma _{\mu \nu }=\underset{\varepsilon \rightarrow 0}{\lim }\left[
K_{\mu \nu }(\eta =+\varepsilon )-K_{\mu \nu }(\eta =-\varepsilon
)\right]
\end{equation*}%
$\eta $ being the arc length measured along the geodesic orthogonal to $%
\Sigma $. The extrinsic curvature $K_{\mu \nu }$ of $\Sigma $ is
given by
\begin{equation}\label{extrinsic_cuvature}
K_{\alpha \beta }=h_{\alpha }^{\mu }h_{\beta }^{\nu }\nabla _{\mu
}n_{\nu }\, .
\end{equation}%

In the present case, the matching is at a timelike hypersurface
$r=r^{\ast }(\tau)$ (let us call such a surface $\Sigma _{(\tau )}$)
which describes the time evolution of a thin spherical shell where
the phase boundary occurs. Namely, at $r=r^{\ast }$ a gapped
transition occurs (such as superconductivity, or Quantum Hall
state).

\subsection{The phase boundary}

Because of the spherical symmetry, it can be assumed that the
spatial sections of the timelike matching hypersurface $\Sigma
_{(\tau )}$ are isomorphic to the two sphere so that there exist a
coordinates system in which the induced metric $\left.
ds^{2}\right\vert _{\Sigma _{(\tau )}}$ and
$S_{\mu }^{\nu }$ respectively read:%
\begin{eqnarray}
\left. ds^{2}\right\vert _{\Sigma _{(\tau )}} &=&-d\tau
^{2}+r^{2}(\tau
)d\Omega ^{2}  \notag \\
S^{\mu \nu } &=&\sigma (\tau )\left( U^{\mu }U^{\nu }\right) -\zeta
\left(
\tau \right) \left( h^{\mu \nu }+U^{\mu }U^{\nu }\right) ,  \label{0emt0} \\
h^{\mu \nu } &=&g^{\mu \nu }-n^{\mu } n^{\nu }  \notag
\end{eqnarray}%
where $h^{\mu \nu }$ is the metric induced on $\Sigma _{(\tau )}$,
$n^{\mu }$ is the (spacelike) normal to $\Sigma _{(\tau )}$, $\tau $
is the arc length measured along the timelike geodesic belonging to
$\Sigma _{(\tau
)}$, $U^{\mu }$ is the normalized four velocity of $\Sigma _{(\tau )}$, $%
\sigma $ is the surface energy density of $\Sigma _{(\tau )}$,
$\zeta \left( \tau \right) $\ is the surface tension and the
conservation of $S_{\mu
}^{\nu }$ implies%
\begin{equation*}
\partial _{\tau }\sigma =-2(\sigma -\zeta )\frac{\partial _{\tau }r}{r}
\end{equation*}%
with $r(\tau )$ being the proper circumferential radius of the domain wall $%
\Sigma _{(\tau )}$.The requirement of a traceless $S^{\mu \nu }$
(otherwise it would not correspond to the classical description of
boundary gapless
degrees of freedom) leads to%
\begin{equation}
\sigma +2\zeta =0  \label{0co0} \end{equation} which combined with
the above conservation equation implies
\begin{equation}
 \sigma =\frac{\sigma
_{0}}{r^{3}},\qquad \sigma _{0}>0 \label{0cococo0}
\end{equation}%
where $\sigma _{0}$ is an integration constant, with the dimension
of an energy times a length, which depends on the microscopic model
and which will be approximated by a characteristic energy scale of
the gapped phase transition times the length scale of the domain
wall.

To get an intuition on the meaning of $\sigma _{0}$ let us consider the case
in which in the inner core of the neutron star there is a Quantum Hall
phase. In such a case, the only physical meaningful energy scale of the
model is the energy gap $E_{gap}$, thus one can expect that%
\begin{equation*}
\sigma _{0}\approx \left( E_{gap}\sum_{i}^{n_{\max }}d(i)\right) r^{\ast }
\end{equation*}%
where $r^{\ast }$\ is a typical value of the radius of the inner core, the
index $i$ labels the Landau levels\footnote{%
In contrast with the ordinary Quantum Hall devices in which one often can
assume that only one Landau Level is fully filled, in the present case
(because of the huge density inside the inner core of a neutron star) one
should expect that many Landau level may be fully filled.}, $d(i)$ is the
degeneracy of the $i$-th level, and $n_{\max }$ is the last fully filled
Landau level (so that $\sigma _{0}/\left( r^{\ast }E_{gap}\right) $
represents the number of particles living in the Hall phase). In the case of
a supeconductivity/superfluidity phase inside the inner core also one should
expect a close relation between $\sigma _{0}/r^{\ast }$ and the energy gap
times a number of the order of magnitude of the number of particles living
in the gapped phase.

The ``equation of motion of the domain wall", that is the equation
which determines the evolution of $r(\tau )$, can be deduced from
the matching condition (\ref{jun1}). In particular, the important
equation is the angular
components $S_{i}^{i}$ of Eq. (\ref{jun1})%
\begin{equation}
\gamma _{i}^{i}=-8\pi GS_{i}^{i}  \label{0eqo0}
\end{equation}%
One can see that the $\tau \tau $ component of the matching equations
\begin{equation}
\gamma _{\tau }^{\tau }=-8\pi GS_{\tau }^{\tau }  \label{0eqo1}
\end{equation}%
is not independent of the angular ones provided the consistency
condition\cite{Is66} $S^\nu_{\mu|\nu} = 0 \Rightarrow n^\nu
T_{\mu\nu}(\text{ext}) - n^\nu T_{\mu\nu}(\text{int}) =0$ for the
conservation of energy
on $\Sigma$ is taken into account: one can easily see that the $%
\tau \tau $ component of the matching equations only constraints the
discontinuity of the first derivative of the metric function $\nu$.
Note that the pressure and density profiles are assumed to be
static.

The standard procedure to compute $\gamma _{i}^{i}$ is as follows: the unit
normal to the junction surface is%
\begin{equation}
n_{\mu }=\pm \left( -\exp (\nu +\lambda )\partial _{\tau }r,\exp (2\lambda )%
\sqrt{\left( \partial _{\tau }r\right) ^{2}+\exp (-2\lambda )},0,0\right)
\label{1nove1}
\end{equation}%
where the $\pm $ ambiguity depends on whether one considers the
inner or the outer directed unit normal. A straightforward
calculation using (\ref{extrinsic_cuvature}) shows that the angular
components of the extrinsic curvature is
\begin{equation}
K_{i}^{i}=\pm \frac{\sqrt{\left( \partial _{\tau }r\right) ^{2}+\exp
(-2\lambda )}}{r}  \label{3nove3}
\end{equation}%
while the $\tau \tau $-component of the extrinsic curvature is
(\cite{Sato})
\begin{equation}
K_{\tau }^{\tau }=\pm \frac{\nu ^{\prime }\left( \left( \partial _{\tau
}r\right) ^{2}+\exp (-2\lambda )\right) +\left( \partial _{\tau
}^{2}r\right) +\lambda ^{\prime }\left( \partial _{\tau }r\right) ^{2}}{%
\sqrt{\left( \partial _{\tau }r\right) ^{2}+\exp (-2\lambda )}}.
\label{4nove4}
\end{equation}%
Eventually, Eq. (\ref{0eqo0}) reads%
\begin{equation}
\left( K_{i}^{i}(int)-K_{i}^{i}(ext)\right) =\frac{4\pi G\sigma _{0}}{r^{3}}%
\equiv \frac{K}{r^{3}}  \label{00pre00}
\end{equation}%
where there is no sum on the repeated indices in this formula, $%
K_{i}^{i}(int)$ and $K_{i}^{i}(ext)$ stand for the extrinsic curvature
computed on the inner and on the outer side of the junction. This equation
can be written as an ordinary first order equation for $r(\tau )$ bringing
on the right hand side $K_{i}^{i}(ext)$ and then squaring:%
\begin{equation*}
\left( \partial _{\tau }r\right) ^{2} =\left[ \frac{\left( \exp
(-2\lambda _{int})-\exp (-2\lambda _{ext})\right)
}{2K}r^{2}+\frac{K}{2r^{2}}\right] ^{2}-\exp (-2\lambda _{ext})
\label{0pot}
\end{equation*}%
which, using (\ref{0matov}) can be written:
\begin{gather}
\left( \partial _{\tau }r\right) ^{2}+V(r) = 0\ ,  \label{0pot0} \\
V(r) \equiv -\left( \frac{2Gm_{ext}(r)}{r}- 1 +\left[ \frac{G r}{K}
(m_{ext}(r)- m_{int}(r)) - \frac{K}{2 r^2} \right]^{2}\right) ,
\label{1pot}
\end{gather}%
where $\lambda _{int}$ and $\lambda _{ext}$ stand for the metric
functions of the inner and of the outer side of the junction and
$m_{ext}(r)$ is the mass function (\ref{0matov}) of the exterior
part. It is worth to note an interesting feature of the present
description: the dynamics of $r(\tau )$ given by equation
(\ref{0pot0}) only depends on the metric function $\exp (-2\lambda
)$ which in turns only depends on the density (see Eq.
(\ref{0matov})).

The pressure inside and outside of the shell is given by the
Tolman-Volkov-Oppenheimer-Snyder equations
\begin{eqnarray}
\partial _{r}P &=&-\left( \rho +P\right) \left( 1-2G\frac{m(r)}{r}\right) ^{-1}
\frac{\left( m+4\pi r^{3}P\right) }{r^{2}}
\end{eqnarray}%
where $m$ and $\rho$ on each side are functions of $r$. Therefore we
get an equation of the form $\partial_r P_{ext} a(r) + b(r) +
c(r)P_{ext}(r) + d(r)P_{ext}^2(r) =0$ (and similarly for $P_{int}$),
where $a$, $b$, $c$ and $d$ are polynomial in $r$, has a solution
(which may be found numerically) with one constant of integration.
For a consistent solution, we must take into account equation
(\ref{0eqo1}). To see the existence of a solution it is sufficient
to consider the static case at the equilibrium radius $r_{eq}$, in
which the equation reduces to:
\begin{equation}
 \left[ \frac{m(r_{eq})+ 4\pi r_{eq}^3 P(r_{eq}) }{ r_{eq}^2
 \sqrt{1-\frac{2m(r_{eq})}{r_{eq}}}}\right]^{ext}_{int} =-2 \frac{K}{r_{eq}^3}
\end{equation}
where $[\cdots]^{ext}_{int}$ means the jump in this quantity across
the shell. Generally, we can see this as an equation constraining
the two integration constants for $P_{int}$ and $P_{ext}$ in terms
of $K$. It is enough to know that there is a value for the pressure,
but the precise value shall not be needed to solve the stability
problem, since the potential depends explicitly on $\rho$ and not
$P$.

One sees also that the equation of motion (\ref{0pot0}) for the
shell $r(\tau )$
is formally analogous to the equation of motion of a classical point particle where $%
V(r) $ takes the role of an effective potential. A necessary
condition to have a static phase boundary is that $V(r)$ must have a
local minimum. The exact form of the potential depends now on the
functional form of the density profiles of the two phases.

It will be now explicitly shown that in a suitable range of
parameters such a local minimum exist. \footnote{An analogous
situation possibility has been studied \cite{Cao7} in the context of
black hole physics, where the conformal degrees of freedom led to
the existence of a local minimum for the effective potential (In
\cite{BGG87}, where a boundary stress tensor of a cosmological term
is considered, there was no local minimum).}

\section{Density profiles and (non)staticity of the phase boundary}

To find the possible relative minima (if any) of the potential
$V(r)$ one has to write explicitly $\lambda _{int}$ and $\lambda
_{ext}$\ as functions of $r$ and of phenomenological parameters
characterizing the density of the star. At first glance, to do this
one should solve the Tolman-Oppenheimer-Volkoff equations inside and
outside the inner core. This is a very hard task since to do this
one should approximate the equation of state of strongly interacting
quarks, gluons, neutrons and so on at very high densities and
pressures. In fact, this can be avoided: the reason is that the
effective potential only depends on the metric functions $\lambda
_{int}$ and $\lambda _{ext}$ (that is, on $g_{rr}$). Such function
is expressed in terms of the density profile: if one is able to
determine the dependence of the density on the radial coordinate $r$
the effective potential can be written explicitly. Remarkably
enough, in many sound models of neutron stars it is possible to
determine the density profile explicitly even if the knowledge of
the equation of state (that is, the functional relation between
density and pressure) is not perfect (see, for instance, \cite{AC07}
\cite{Ste03} \cite{LP00} and references therein). On the basis of
the above references, it will be here assumed that the density of
the
neutron star has the following form:%
\begin{eqnarray}
\rho _{int}(r) &=&\frac{A_{int}}{4\pi }-\frac{B_{int}}{4\pi }r^{n},\ \ \
r<r(\tau )  \label{depro1} \\
\rho _{ext}(r) &=&\frac{A_{ext}}{4\pi }-\frac{B_{ext}}{4\pi }r^{n},\ \ \
r>r(\tau ),  \label{depro2}
\end{eqnarray}%
where%
\begin{equation*}
n=1,\ \ A_{int},A_{ext}\approx \left( 10^{11}\sim 10^{15}\right) g/cm^{3},\
\ \ B_{int},B_{ext}\approx \left( 10^{11}\sim 10^{15}\right) g/cm^{4}.
\end{equation*}%
Of course, even if it is not explicitly present, the pressure is needed to
determine the values of the above density profiles. The choice of a linear
decreasing is general enough since in the present case the important region
is the one near the boundary phase so that a linear decreasing captures the
main qualitative features (for instance, if one would take a quadratic
decreasing (that is, $n=2$ in Eqs. (\ref{depro1}) and (\ref{depro2})) as in
\cite{LP00} the main conclusions of the present paper would not change). On
the other hand, interesting phenomenological conclusions can be reached in
the cases in which further terms are added in the density profile. Such
consequences will be discussed in a moment. Therefore, one can assume that
both in the inner core ($r<r(\tau )$) and outside the core ($\ r>r(\tau )$)
the density decreases linearly. With the above choice of the parameters the
effective potential (\ref{1pot}) reads%
\begin{align}
V(r) &=-\bigg( 2G\left( \frac{A_{ext}}{3}r^{2}-\frac{B_{ext}}{4}%
r^{3}\right) -1  +\bigg.  \notag \\
&\quad+\bigg. \frac{1}{r^{4}}\left[ G\left( \frac{A_{ext}-A_{int}}{3K}%
\right) r^{6}-G\left( \frac{B_{ext}-B_{int}}{4K}\right) r^{7}+\frac{K}{2}%
\right] ^{2}\bigg)  \label{11pot11}\, .
\end{align}%
To proceed, let us adopt the geometric unit of measure in which any
quantity is expressed in powers of $cm$ (and the Newton constant is
$1$):
\begin{equation*}
1\ MeV \approx 10^{-55}cm,\qquad G=1 \, .
\end{equation*}
Typical values are:
\begin{eqnarray*}
A_{ext} &\approx& 10^{-16}cm^{-2},\quad B_{ext}\approx
10^{-19}cm^{-3}\, , \qquad K\approx 10^{-29}cm^{2},
\end{eqnarray*}%
having assumed%
\begin{equation*}
E_{gap}\approx MeV,\ \ \ N\approx 10^{25},\ \ \ r^{\ast }\approx cm
\end{equation*}%
where $N$ is the order of magnitude of the number of particles
living in the inner phase and $r^{\ast }$ is the typical radius of
the inner core.

\subsection{Stable range}

From a numerical analysis, it can be seen that in order for there to
be a minimum of the potential we must have the following relations
\begin{equation*}
\Delta A\Delta B>0,
\end{equation*}%
a further condition is that
\begin{equation}
\left( \partial _{\tau }r\right) ^{2}+V(r) = 0 \quad \Rightarrow
\quad
 V(r) \leq 0.  \label{depro3}
\end{equation}%
When the above condition is not fulfilled, the Israel equations
cannot be satisfied at all (the matching would be impossible). The
numerical graphs also suggest that for a given $\Delta B$ there is a
minimum value of $\Delta A$ below which solutions do not exist. For
a core of radius $\approx 1 \ cm$ this minimum is $\Delta A_{min}
\approx \Delta B/2$ as shown in graph \ref{Stable_Range}. More
generally, one can check the following:
\begin{equation}
 \Delta M = \frac{\Delta A}{3}r^3 - \frac{ \Delta B}{4} r^4 \geq
 Kr^2\, .
\end{equation}
This is natural (by analogy with special relativity) since $K r^2$
is the rest mass of the shell and $\Delta m$ is a measure of the
relativistic mass. If this condition is not fulfilled, there is
either no minimum of the potential, or the minimum is positive,
meaning no stable solution (see figure 2 ).

\begin{figure}[tbh]
\centering
\includegraphics[width=4.5in]{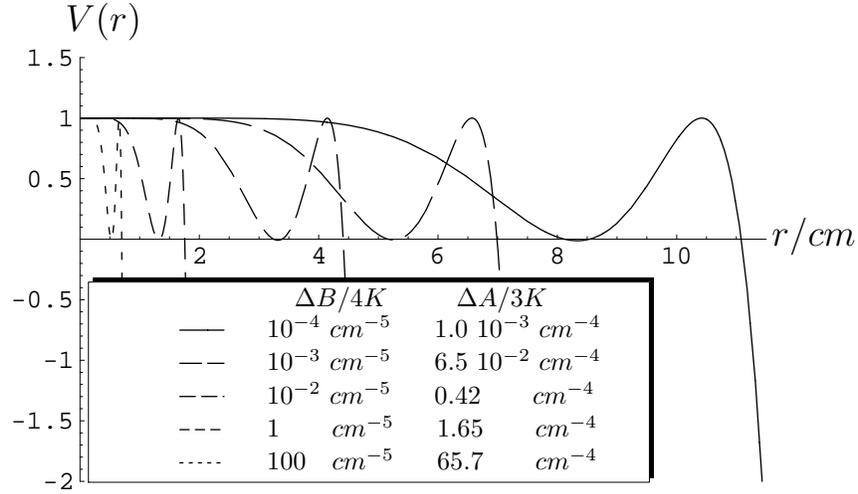}
\caption{{\small In this figure there is a graph of the effective
potential for varying $ 10^{-4} cm^{-5} \leq \Delta B/ 4K \leq 100
cm^{-5}$ and suitable values of $\Delta A/ 3K$ such that the
potential has a minimum for small negative $V$. Stable solutions
with radius of order $1-10$ $cm$
 are shown to exist.}} \label{Range of r}
\end{figure}

\begin{figure}[tbh]
\centering
\includegraphics[width=5in]{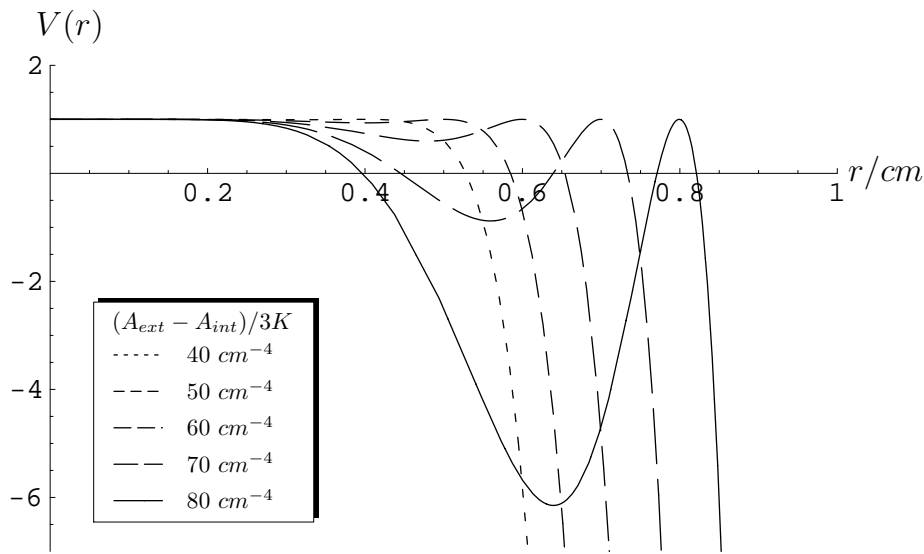}
\caption{{\small In this figure there is a graph of the effective
potential for fixed $\Delta B/ 4K = 100 cm^{-5}$ and varying values
of $\Delta A/ 3K$ (see plot legend above). Stable solutions exist
for $\Delta A/ 3K$ greater than approximately $65 cm^{-4}$.
}}\label{Stable_Range}
\end{figure}

In the range of parameters we focus on (see graph \ref{Range of r})
a stable core exists: the equilibrium radius $r_{eq}$
and the frequency $\omega _{pul}$ are%
\begin{align}
 r_{eq}\approx& 0.53\ cm,\qquad \left. \partial _{r}^{2}V
 \right\vert_{r_{\min }}=  146 cm^{-2}\, , \label{sessea}
 \\
  r_{eq}\approx& 1.32\ cm,\qquad \left. \partial _{r}^{2}V
 \right\vert_{r_{\min }}=  23.1 cm^{-2}\,\notag
 \\
  r_{eq}\approx& 3.32\ cm,\qquad \left. \partial _{r}^{2}V
 \right\vert_{r_{\min }}=  3.67 cm^{-2}\,\notag
 \\
  r_{eq}\approx& 5.26\ cm,\qquad \left. \partial _{r}^{2}V
 \right\vert_{r_{\min }}=  1.46 cm^{-2}\,\notag
 \\
  r_{eq}\approx& 8.34\ cm,\qquad \left. \partial _{r}^{2}V
 \right\vert_{r_{\min }}=  0.582 cm^{-2}\,\notag
\end{align}%
So the frequencies are of order of GHz.

The above frequency of pulsation would correspond to a periodic
motion of $r(\tau )$ between a maximum and a minimum radius
oscillating around the above
equilibrium radius\footnote{%
The actual possibility that such an exact pulsation mode could be
realized in practice in a typical neutron star will be discussed in
the next section.}. The equilibrium radius considered here is of the
order of centimeters (which is in good agreement with the QFT
estimates in the case of the Quantum Hall Effects \cite{Iw05}
\cite{IM05}; such a value for the radius of the inner core could
also be
compatible with the presence of a superconductivity/superfuidity phase \cite%
{Pet92} \cite{Ra99}).

\subsection{Nonstatic range}

The numerical analysis shows that if \ $\Delta A$ and $\Delta B$ do
not have the same sign the configuration is unstable: that is, the
effective potential (\ref{11pot11}) has only one local maxima and
the phase boundary will tend to move. It is important to stress that
this result is a purely general relativistic effect related to the
conformal boundary degrees of freedom. Such an effect would be
overlooked if one neglects the possible dynamics of the phase
boundary. If one construct a neutron star by matching the inner core
and the outer part at a fixed radius $r^{\ast }=const$ (neglecting
its possible dependence on $\tau $) and without the inclusion of the
boundary degrees of freedom then it is not possible to disclose the
effective potential.

\subsection{Multiple equilibria?}

As it has been already discussed, a density which decreases linearly with
the radius is reasonable and general. On the other hand, it may be well
possible (see \cite{AC07} \cite{Ste03} and references therein) that the
density profile is corrected by further terms as for instance%
\begin{eqnarray}
\rho _{int}(r) &=&\frac{A_{int}}{4\pi }-\frac{B_{int}}{4\pi }%
r+C_{int}r^{n},\ \ \ r<r(\tau )  \label{fuffuo1} \\
\rho _{ext}(r) &=&\frac{A_{ext}}{4\pi }-\frac{B_{ext}}{4\pi }%
r+D_{ext}r^{m},\ \ \ r>r(\tau ),  \label{fuffuo2} \\
n,m &>&1.  \label{fuffuo3}
\end{eqnarray}%
At a first glance, one could expect that near the phase boundary
only the first two terms in Eqs. (\ref{fuffuo1}) and (\ref{fuffuo2})
are important. However, if due to the extreme conditions inside a
neutron star, such further terms have to be taken into account, then
interesting phenomenological scenarios open up. In these cases, the
effective potential may have multiple local minima and the valleys
would be generically asymmetric: namely the valley corresponding to
the larger radius of equilibrium could be more deep than the valley
corresponding to the smaller radius of equilibrium. In principle, a
phase transition could occur in which the phase boundary jumps from
one valley to the other.

\section{Astrophysical Implication}

From the explicit form of the effective potential (\ref{11pot11}) it
is clear that, in the range of parameters in which the effective
potential has a local minimum, it is consistent to assume a static
phase boundary. In this range of parameters also small oscillations
of the phase boundary could be possible. However it is difficult to
say what would be the sources of dissipation which would damp the
oscillations.

It is a well known fact since the Migdal paper \cite{Mi59} that the
presence of a superfluid phase inside a neutron star affects its
moment of intertia. In the range of parameters in which the phase
boundary tends to move one phase tends to swallow the other. In the
case in which the inner phase expands, there is a corresponding
decrease of the moment of intertia. Assuming that our results also
hold for the rotating case (which is a reasonable assumption at
least in the slow rotating case \cite{Ste03} \cite{FG07}) this would
lead to interesting observable effects: namely a sudden small
increase of the angular velocity because of the angular momentum
conservation. Analogously, in the case in which the inner phase
shrinks, there is a corresponding decrease of the angular velocity.
Indeed both types of phenomena are of interest in astrophysics.

\section{ Conclusions and perspectives}

We have studied a model of a neutron star with an inner core which
undergoes a phase transition. It has been shown that, even assuming
mechanical equilibrium, it is not always consistent to assume a
static phase boundary once conformal boundary degrees of freedom are
taken into account.

In order to study the staticity of such a phase boundary one must
take into account the general relativistic effects of these boundary
degrees of freedom by including a nontrivial stress tensor on the
junction. To the best of the authors knowledge, such a consistency
analysis with conformal boundary degrees of freedom has not been
performed previously. The astrophysical consequences of the
non-static regime, related to sudden changes of the moment of
inertia of the star, are worth further investigating.

\section*{Acknowledgments}

We want to thank R. Troncoso and J. Zanelli for important
suggestions and useful criticism. We also want to thank A. Anabalon,
H. Maeda, J. Oliva for many useful discussions. The work of F. C.,
A.G. and S.W. has been partially supported by
Proy. FONDECYT N%
${{}^\circ}$%
3070055, 3070057 and 3060016. This work was funded by an
institutional grants to CECS of the Millennium Science Initiative,
Chile, and Fundaci\`{o}n Andes, and also
benefits from the generous support to CECS by Empresas CMPC. 

\end{document}